\documentclass{article}
\usepackage{spconf,amsmath,graphicx}
\usepackage{url}
\usepackage{xcolor}
\usepackage{booktabs}
\usepackage{siunitx}
\usepackage{multirow, makecell}
\usepackage{bbold}
\usepackage{pifont}
\newcommand{\cmark}{\ding{51}}%
\newcommand{\xmark}{\ding{53}}%
\usepackage{enumitem}
\usepackage[normalem]{ulem}

\title{MULTI-SPEAKER AND WIDE-BAND SIMULATED CONVERSATIONS \\ AS TRAINING DATA FOR END-TO-END NEURAL DIARIZATION}

\name{\begin{tabular}{c}Federico Landini$^1$, Mireia Diez$^1$, Alicia Lozano-Diez$^2$, Luk\'a\v{s} Burget$^1$\thanks{The work was supported by Czech Ministry of Interior project No. VJ01010108 "ROZKAZ", Czech National Science Foundation (GACR) project NEUREM3 No. 19-26934X, Horizon 2020 Marie Sklodowska-Curie grant ESPERANTO, No. 101007666 and grant PID2021-125943OB-I00, MCIN /AEI /10.13039/501100011033 / FEDER, UE. Computing on IT4I supercomputer was supported the Ministry of Education, Youth and Sports of the Czech Republic through the e-INFRA CZ (ID:90140).}\end{tabular}}

\address{$^1$Brno University of Technology, Faculty of Information Technology, Speech@FIT, Czechia \\
        $^2$AUDIAS (Audio, Data Intelligence and Speech), Universidad Aut\'onoma de Madrid, Spain \\
        \textit{\{landini,mireia,burget\}@fit.vutbr.cz, alicia.lozano@uam.es}}

\begin{document}
\ninept
\maketitle

\begin{abstract}
End-to-end diarization presents an attractive alternative to standard cascaded diarization systems because a single system can handle all aspects of the task at once. Many flavors of end-to-end models have been proposed but all of them require (so far non-existing) large amounts of annotated data for training. The compromise solution consists in generating synthetic data and the recently proposed simulated conversations (SC) have shown remarkable improvements over the original simulated mixtures (SM). In this work, we create SC with multiple speakers per conversation and show that they allow for substantially better performance than SM, also reducing the dependence on a fine-tuning stage. We also create SC with wide-band public audio sources and present an analysis on several evaluation sets. Together with this publication, we release the recipes for generating such data and models trained on public sets as well as the implementation to efficiently handle multiple speakers per conversation and an auxiliary voice activity detection loss.
\end{abstract}
\vspace{-0.15cm}

\vspace{-0.3cm}

\section{Introduction}
\label{sec:intro}
\vspace{-0.3cm}

End-to-end neural diarization (EEND) is a popular alternative to the traditional cascaded speaker diarization system composed of different submodules, i.e. voice activity detection (VAD), uniform segmentation, speaker embedding extraction, clustering and overlapped speech detection and handling. Instead of having separate models trained independently, EEND formulates the speaker diarization problem as a per-speaker-per-time-frame binary classification problem where a permutation-free objective is used to minimize the speech activity error for
all speakers. This has the advantage of training the model directly for the task of interest; but unlike in the cascaded paradigm, end-to-end models require large amounts of data with diarization annotations. Given that currently available data with manual annotations for diarization are scarce, the usual strategy is to create audios where segments of different recordings are combined and VAD annotations define the ground truth annotations. 

When EEND was proposed, Fujita et al. defined a procedure to generate training data in the form of simulated mixtures (SM)~\cite{fujita2019enda} Algorithm 1. The method selects the number of speakers that the mixture will have and for each speaker, speech segments are selected from a pool. Then, pauses are introduced between the segments of a speaker. This is repeated for each speaker independently, producing one channel per speaker. Finally, the channels are summed to produce the final mixture. In order to enrich the mixtures, the speech from each speaker is reverberated and background noise is added to the final mixture.
The main disadvantage of this approach is that the pauses between the turns of each speaker (and resulting overlaps) are defined independently and this does not resemble real conversations. In order to tackle this problem, approaches that consider the interaction between speakers were proposed~\cite{landini2022simulated,yamashita2022improving}. Instead of treating speakers independently, simulated conversations (SC)~\cite{landini2022simulated} Algorithm 1, are defined by interleaving speech segments and determining pauses or overlaps between them with probabilities and lengths drawn from distributions estimated on real interactions. SC can also be optionally enriched with reverberation and background noises. It has been shown \cite{landini2022simulated} that SC allow for better performing EEND models on 2-speaker telephone conversations and reduce the dependence on a fine-tuning stage, i.e. after training the model with SC, it is further trained using a small development set with low learning rate.

Most works with end-to-end models so far have focused on the telephony scenario where large amounts of data are available and where each speaker is recorded in a separate channel. This allows to create diarization segment annotations by simply running a VAD system on each channel. This scenario presents certain advantages such as (usually) one speaker per channel, conversational speech and relatively similar channel characteristics across recordings. This last aspect allows for a relatively easy amalgamation of different recordings when creating synthetic data where the channel differences are less relevant than the speaker ones and, thus, the model can focus on learning to separate speakers rather than channels. 

Contrarily, wide-band data can have more variability in terms of recording devices, settings and types of interactions. However, it is rare to have a single speaker per recording, even less in a conversational scenario. Thus, applying the same techniques for creating synthetic data becomes more challenging since the mismatch between (artificial) training and (authentic) test data can become much larger.
There have been efforts~\cite{leung2021end,liu2021end,kinoshita2021integrating,maiti2021end} in generating training data for EEND using wide-band recordings from LibriSpeech~\cite{panayotov2015librispeech} mostly based on SM. In this work, we aim to extend the usage of SC to wide-band data while exploring different available corpora. For the sake of keeping the focus on the data sources for SC, we perform the analysis under the scenario of two speakers per conversation. 

In order to analyze the advantages of SC for more than two speakers, we focus on the already studied telephony case where SC has been shown to be more realistic than SM~\cite{landini2022simulated}. Furthermore, a modification in the loss is introduced in order to be able to train models with several speakers efficiently and an auxiliary loss is added to improve the model's VAD capabilities.

\section{Wide-band Simulated Conversations}
\vspace{-0.1cm}

Although there are nowadays numerous wide-band collections with thousands of hours of speech, they contain normally more than one speaker in the same channel. This makes it impossible to use the same strategy as with telephone for deriving the speaker turns. Instead, it is necessary to have some kind of segmentation already available that contains information about the speakers. Next, we describe the freely available datasets that we used for this work\footnote{It should be noted that GigaSpeech~\cite{chen2021gigaspeech} was considered but, at the time of writing, segments do not have speaker labels so it is not usable for our purposes.}.

\begin{itemize}[label={$\bullet$}, topsep=0pt, itemsep=0pt, leftmargin=10pt]
    \item LibriSpeech~\cite{panayotov2015librispeech} consists of 1000 hours of read English speech from almost 2500 speakers. Each recording is expected to contain speech of a single speaker; thus, the original strategy of running VAD to obtain segmentation is possible. Recordings are of good quality and without background noises meaning that channel characteristics are relatively similar across recordings. However, all speech is read and not conversational.
    \item VoxCeleb2~\cite{chung2018voxceleb2} consists of more than 2400 hours of recordings from more than 6000 speakers speaking mostly English. Originally prepared as a training set for training speaker recognition systems, the recordings are partially annotated. This means that for the speakers of interest, some of their segments are identified. Thus, it is possible to derive speech segments for a given speaker without the need for any VAD system. At the same time, these annotations are automatically generated, possibly introducing errors such as including small excerpts from other speakers. Furthermore, the speech is collected ``in the wild'' so the recordings can have different noises. We observed better results if recordings with a poor signal-to-noise ratio (SNR) were filtered out.
    \item VoxPopuli~\cite{wang-etal-2021-voxpopuli} consists of recordings from the European Parliament in different languages. For this work, we used the subset for which transcriptions exist. Each recording is expected to contain speech from a single speaker and the transcription timestamps are used to derive the segmentation used for SC. Since the annotations were automatically derived, in some rare cases there are short excerpts from other speakers. This subset contains approximately 2700 hours from 2600 speakers. The recordings correspond to speakers' turns during plenary sessions which are monologues. Therefore, this corpus presents speech that is not read nor conversational but has exclusive turns and spontaneous speech. Recordings are normally of good quality and without background noises.
\end{itemize}

As in the telephony scenario, SC were augmented with background noises  from the MUSAN collection~\cite{snyder2015musan} scaled with an SNR selected randomly from \{5, 10, 15, 20\} dB. Room impulse responses and leveling of relative energy between speakers were evaluated as mentioned in~\cite{landini2022simulated} but the performance was about the same so we present results only when adding background noises.

\section{Simulated Conversations with Several Speakers and Model Adjustments}
\label{sc_multispk_vadloss}
\vspace{-0.1cm}

SC are more realistic than SM in terms of the percentages of silences and overlap in the produced recordings\cite{landini2022simulated}. Given that each speaker is modeled independently in SM, the percentages of overlapped speech are usually much larger than in real conversations. This is exacerbated the more speakers per mixture are used. On the contrary, SC contain proportions of silence and overlap that still resemble those seen in real data. We compare the performance of SM and SC with recordings with more than two speakers.

We carry out our analysis using the self-attention EEND with encoder-decoder attractors (EEND-EDA)~\cite{horiguchi2020end}. One of its known limitations is the inability to handle more speakers per utterance than those seen during training, i.e. if the model is trained with recordings that contain up to 4 speakers, it will not perform well for recordings with more than 4. At the same time, one of the limitations in the setup originally proposed in~\cite{horiguchi2020end} is that with the permutation invariant training (PIT) scheme, naively calculating all permutations becomes prohibitive in practice for more than 4 or 5 speakers per sequence. Faster alternatives have been studied~\cite{lin2020optimal} and in this work, we use PIT to find the best assignment between speakers and reference labels in polynomial time using the Hungarian algorithm\footnote{In fact, we use scipy.optimize.linear\_sum\_assignment  which implements the Jonker-Volgenant variant.}. This allows us to be able to train the models on utterances with more speakers without increasing the computational cost considerably.

In previous experiments with SC~\cite{landini2022simulated} it was observed that the outputs of the model have relatively high missed and false alarm speech. For this reason, and following a similar idea to that proposed in~\cite{zheng2022cuhk}, we introduced the auxiliary VAD loss in Eq.~\ref{eq:vad_loss} which reinforces per frame speech/non-speech classification.
\vspace{-0.2cm}
\begin{equation}
L_{VAD} = -\frac{1}{F} \sum_f^F s_f \log (p(sil_f)) + (1-s_f) \log (1-p(sil_f)) 
\label{eq:vad_loss}
\end{equation}
where 
$p(sil_f) = \prod_s (1-y_f^s)$
represents the probability of silence given by the model for frame $f$,
$s_f = \mathbb{1}[(\sum_s t_f^s) = 0]$
represents the silence label for frame $f$, 
where $t_f^s$ and $y_f^s$ are the label and model probability for activity of speaker $s$ at frame $f$ respectively.
This loss is combined with the diarization and attractor existence losses in the following fashion
$L = L_{diarization} + L_{attractors} + \alpha L_{VAD}$.

\section{Experimental setup}
\vspace{-0.1cm}
\subsection{Diarization model}
\label{diarization_model}
\vspace{-0.1cm}
All experiments were performed using the EEND-EDA for showing superior performance in previous works~\cite{horiguchi2020end}. The architecture used was exactly the same as that described in~\cite{horiguchi2020end}\footnote{We refer the reader to~\cite{horiguchi2020end} for a scheme of the model.}. For the sake of making the code more efficient, we used our PyTorch implementation\footnote{https://github.com/BUTSpeechFIT/EEND}. 15 consecutive 23-dimensional log-scaled Mel-filterbanks (computed over 25\,ms every 10\,ms) are stacked to produce 345-dimensional features every 100ms. These are transformed by 4 self-attention encoder blocks (with 4 attention heads each) into a sequence of 256-dimensional embeddings. These are then shuffled in time and fed into the LSTM-based encoder-decoder module that decodes attractors, which are deemed as valid if their existence probability is above a certain threshold. A binary linear classifier is used to obtain speech activity probabilities for each speaker (represented by a valid attractor) at each time step (represented by an embedding). 

The training scheme consists in training the model first on synthetic training data and then performing fine-tuning using a small development set of real data of the same domain as the test set. Usually, in the experiments with more than two speakers, a model initially trained on synthetic data with two speakers per recording is adapted to a synthetic set with more speakers and finally fine-tuned to a development set. The initial training is run for 100 epochs, the adaptation is run for 100 epochs on smaller sets or 25 epochs on (approximately four times) larger sets. The fine-tuning step is run for 100 epochs.

During inference time, classifiers' outputs are thresholded at 0.5 to determine speech activities. 
Each training was run on a single GPU. The batch size was set to 64 or 32 with 100000 or 200000 minibatch updates of warm-up respectively. Following~\cite{horiguchi2020end}, the Adam optimizer~\cite{kingma2014adam} was used and scheduled with noam~\cite{vaswani2017attention} reaching a maximum learning rate of $10^{-3}$. For fine-tuning on a development set, the Adam optimizer was used with learning rate $10^{-5}$. For the inference as well as for obtaining the model from which to fine-tune or adapt, the models from the last 10 epochs were averaged. Unless otherwise specified, during training, adaptation and fine-tuning, batches were formed by sequences of 500 Mel-filterbank outputs, corresponding to 50\,s. During inference, the full recordings are fed to the network one at a time. 
Diarization performance is evaluated in terms of diarization error rate (DER) as defined by NIST~\cite{NISTRT}.
For evaluation sets where a forgiveness collar is used when calculating DER, a median filter with window 11 is applied as post-processing over the network's output. If the forgiveness collar is 0\,s, no filtering is applied as this provides better definition in the output.

\subsection{Evaluation sets for wide-band experiments}
\vspace{-0.1cm}
Different collections were used for evaluation. For the wide-band scenario, only two-speaker recordings were used. We evaluate results on the four domains from the Third DIHARD Challenge~\cite{ryant2020third} that satisfy such condition: CTS, Clinical, Maptask and Sociolinguistic (lab). More information can be found in~\cite{ryant2020third} and in Table~\ref{tab:evaluation_datasets}.

\begin{table}[t]
    \caption{Evaluation sets for wide-band scenario. For AMI, numbers do not correspond to dev but to train set.}
    \label{tab:evaluation_datasets}
    \setlength{\tabcolsep}{3pt} 
    \centering
    \begin{tabular}{llcccc}
    \toprule
    \multirow{2}{*}{Name} & \multirow{2}{*}{Type} & \multicolumn{2}{c}{\#files} & \multicolumn{2}{c}{Avg. length (s.)} \\
    & & dev & eval & dev & eval \\
    \midrule
    CTS & telephone conversations & 61 & 61 & 600 & 600 \\
    Clinical & interviews with children & 48 & 51 & $\approx$300 & $\approx$300 \\
    Maptask & fast-paced interactions & 23 & 19 & $\approx$400 & $\approx$400 \\
    Socio lab & interviews with adults & 16 & 12 & $\approx$600 & $\approx$600 \\
    AMI & meetings (2-spk) & 804 & 93 & 1190 & 1137 \\
    VoxConverse & broadcast (2-spk) & 44 & 31 & 280 & 520 \\
    
    \bottomrule
  \end{tabular}
  \vspace*{-0.2cm}
\end{table}

Another domain of interest for diarization is meetings. However, there are no datasets in a meeting-like scenario with only two speakers. Given that many end-to-end diarization works still focus on the two-speaker scenario, we considered of relevance to derive from AMI meetings~\cite{carletta2005ami} all possible subsets of two speakers for each conversation and make it of public access\footnote{https://github.com/BUTSpeechFIT/AMI\_2speaker\_subset}. For each recording, all pairs of speakers were drawn and, for each pair, all speech where another speaker spoke was removed from the waveforms using reference diarization annotations of the ``only words'' setup described in~\cite{landini2022bayesian}\footnote{https://github.com/BUTSpeechFIT/AMI-diarization-setup}. Then, for each original conversation with four speakers, six \textit{conversations} of two speakers were created. We evaluate results on Mix-Headset audios and the beamformed microphone array N1, where BeamformIt~\cite{Anguera07Beamforming} is applied using the specific AMI setup.

Finally, to add more diversity, recordings with two speakers from VoxConverse~\cite{chung2020spot} were used as these come from varied broadcast sources and present different characteristics from those covered in previously mentioned sets. Following its corresponding evaluation setup, a forgiveness collar of 0.25\,s was used to compute DER while all the sets mentioned above were scored with forgiveness collar 0\,s. In all cases, all speech (including overlap) was evaluated.

\subsection{Evaluation sets for multi-speaker experiments}
\vspace{-0.1cm}

To evaluate the performance when training EEND-EDA with more than two speakers, the 2000 NIST Speaker Recognition Evaluation~\cite{przybocki2001nist} dataset, usually referred as ``Callhome''~\cite{NISTSRE2000evalplan} was used.
We report results using the standard Callhome partition\footnote{https://github.com/BUTSpeechFIT/CALLHOME\_sublists}. We will refer to the parts as CH1 and CH2. 
The amounts of recordings for CH1/CH2 per amount of speakers in the recording are: with 2 speakers 155/148, with 3 61/74, with 4 23/20, with 5 5/5, with 6 3/3, with 7 2/0.
Results on Callhome consider all speech (including overlap segments) for evaluation with a forgiveness collar of 0.25\,s. Fine-tuning was done using CH1, and evaluation on CH2.

\begin{figure*}
    \centering
    \includegraphics[height=4.5cm]{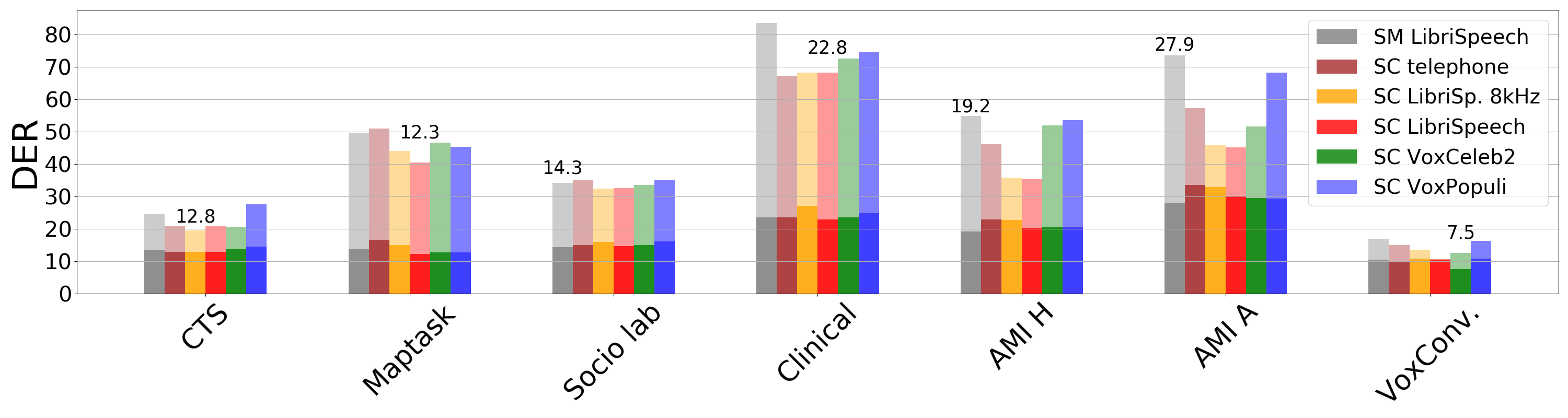}
    \vspace{-0.4cm}
    \caption{Comparison of type of recordings used to generate SC (or SM): LibriSpeech, telephone (Switchboard and SRE), VoxCeleb2 and VoxPopuli. Results on the evaluation/test sets of different subsets of DIHARD3, 2-speaker AMI mix-headset (H) and beamformed array (A) and 2-speaker recordings from VoxConverse. Light shade colors show results before fine-tuning and darker colors refer to results after fine-tuning to a matching development set. Fine-tuning for AMI was for 20 epochs on the train set. Numbers mark the best result among bars.}
    \label{fig:audio_sources} 
\end{figure*}

\vspace{-0.1cm}

\section{Results}
\vspace{-0.2cm}

\subsection{Wide-band experiments}
\vspace{-0.1cm}

One of the main aspects when generating synthetic training data is the choice of recordings to use. Figure~\ref{fig:audio_sources} presents a comparison of SM and SC using different sets, namely telephone conversations, LibriSpeech downsampled to 8\,kHz (mimicking a telephone channel scenario), LibriSpeech, VoxCeleb2 and VoxPopuli. 16\,kHz evaluation data were downsampled to run the inference with 8\,kHz models and 8\,kHz data were upsampled (with empty upper spectrum) to run the inference with 16\,kHz models.

Even though some clear differences exist before fine-tuning, such as SC LibriSpeech performing the best for Maptask and both AMI sets, differences are reduced after fine-tuning. As observed in~\cite{landini2022simulated}, differences of up to 0.5\,DER can easily be attributed to chance; thus, here there are no statistically significant ``winners'' in most cases.
Some of the patterns worth mentioning are that 8\,kHz models perform the worst for Maptask and the AMI sets, suggesting that information on higher frequencies is more relevant in these sets; and that using VoxCeleb2 recordings allows for better performance in VoxConverse, probably due to the similarities between both sets, formed by diverse recordings collected from YouTube.

Unlike the effect seen in the telephony scenario where SC are clearly superior and the need for fine-tuning is reduced substantially\footnote{Results on CTS when training with SM telephone are 4.9\% and 5.2\% relatively worse than when training with SC telephone before and after fine-tuning respectively.}~\cite{landini2022simulated}, when working with different wide-band datasets, fine-tuning still plays a major role and even SM allow for similar performance. Unfortunately, this shows that the main challenge in wide-band scenarios is, until now, not the realism or naturalness of the synthetic data in terms of turns but rather differences in the channel between source and test data, quality of speaker annotation, and conversationality or a combination of all these. 

The mechanism proposed for generating SC makes use of statistics about pauses and overlaps observed in real conversations. We explored different sets for extracting such statistics: CTS, Maptask, Socio lab, Clinical, all the aforementioned together (in the core setup, to have them equally represented), AMI and VoxConverse generating SC using audios from LibriSpeech but there were no significant differences.
While these findings do not imply that it is not possible to produce synthetic data of superior quality that will permit better performance without the need for fine-tuning, we intend to share our results with the community simply to shed light on the directions that have already been explored. There are still other aspects to consider such as improving data augmentation mechanisms like using more realistic background noises, reverberation or loudness levels that match the application scenarios. For instance, CTS, Maptask and AMI H have speech recorded with close-talk microphones, and Socio lab, Clinical and AMI A were recorded with far-field microphones. These aspects were not particularly considered in our analysis but they might have a strong influence\footnote{We generated SC from specific domains, i.e., from Maptask recordings only, and evaluated the performance on the same domain but the performance was considerably inferior; probably due to the limited amount of speakers.}. 

Finally, we hope that by sharing recipes\footnote{https://github.com/BUTSpeechFIT/EEND\_dataprep} for generating training data using public datasets and the models trained on publicly available data, the community will have easier access to baselines that otherwise require expensive computations.

\begin{table}[t]
    \caption{DER comparison on CH2. SM 2\,spk and SC 2\,spk contain $\approx$2500\,h, \textbf{2}-4\,spk contains $\approx$2500\,h (1250\,h recordings with 2 speakers, 625\,h with 3 and 4 each), 2-\textbf{4}\,spk contains $\approx$2500\,h (1250\,h recordings with 4 speakers, 625\,h with 2 and 3 each). 2-7\,spk contains $\approx$2500\,h with recordings following the proportions of speakers seen in CH1. SM 1-4\,spk contains $\approx$15500\,h (100k mixtures of each amount of speakers) and SC contains $\approx$10000\,h (2500\,h of each amount of speakers). Sequences of 200\,s were used for fine-tuning.}
    \label{tab:multispeaker}
    \setlength{\tabcolsep}{5pt} 
    \centering
    \begin{tabular}{@{}
                  l |
                  S[table-format=2.2] |
                  S[table-format=2.2]
                  S[table-format=2.2]
                  S[table-format=2.2] 
                  S[table-format=2.2]
                  S[table-format=2.2] 
                  @{}}
    \toprule
    System & All & \multicolumn{1}{c}{2-spk} & \multicolumn{1}{c}{3-spk} & \multicolumn{1}{c}{4-spk} & \multicolumn{1}{c}{5-spk} & \multicolumn{1}{c}{6-spk} \\
    \midrule
    SM 2 spk & 28.67 & 16.85 & 27.46 & 40.4 & 52.94 & 50.51 \\
    \hspace{0.2cm}+ SM 1-4 spk & 26.14 & 16.28 & 24.67 & 34 & 50.15 & 53.24 \\
    \hspace{1cm}+ CH1 & 17.45 & 8.38 & 16.14 & 24.26 & 36.75 & 46.79 \\ \cline{2-7}
    
    SM 1-4 spk & 27 & 16.1 & 25.44 & 37.56 & 46.52 & 54.92 \\
    \hspace{1cm}+ CH1 & 25.78 & 13.92 & 24.77 & 34.62 & 43.77 & 66.37 \\
    \midrule
    
    SC 2 spk & 20.86 & 8.48 & 21.07 & 29.56 & 45.61 & 49.2 \\
    \hspace{0.2cm}+ SC 1-4 spk & 16.18 & 8.95 & 13.78 & 21.22 & 37.35 & 46.32 \\
    \hspace{1cm}+ CH1 & 16.07 & 10.03 & 14.35 & 19.3 & 30.67 & 46.94 \\ 
    
    \hspace{0.2cm}+ SC \textbf{2}-4 spk & 17.52 & 9.07 & 15.11 & 23.3 & 38.55 & 54.03 \\
    
    \hspace{0.2cm}+ SC 2-\textbf{4} spk & 17.09 & 9.55 & 14.69 & 22.5 & 34.96 & 51.46 \\
    
    \hspace{0.2cm}+ SC 2-7 spk & 17.49 & 9.16 & 15.43 & 24.18 & 39.17 & 45.41 \\
    \cline{2-7}
    
    SC 1-4 spk & 19.9 & 10.2 & 17.79 & 26.49 & 42.67 & 58.09 \\ 
    \hspace{1cm}+ CH1 & 21.24 & 15.45 & 18.89 & 25.17 & 38.16 & 49.54 \\ 
    
    \bottomrule
  \end{tabular}
  \vspace*{-0.2cm}
\end{table}

\vspace{-0.2cm}

\subsection{Multi-speaker experiments}
\label{multispeaker_experiments}
\vspace{-0.2cm}

Table ~\ref{tab:multispeaker} presents a comparison when different sets are used for training, adapting (to more speakers) and fine-tuning (to CH1). The first rows show the performance when using SM. When following the same scheme with SC, the model trained on 2 speakers is already considerably better but adapting to more speakers pushes the performance further. The fine-tuning step provides only small gains, reducing the dependence on this step. Analogously, if the model is directly trained on the 1-4 sets, SC allow for much better performance than SM. However, this set is so biased towards fewer speakers that the model cannot learn from the fine-tuning.

We explored using other SC multi-speaker sets with different proportions of recordings. Training on a set with higher proportion of recordings with 4 than 2 speakers (+SC 2-\textbf{4}\,spk) is beneficial since the resulting model can deal better with recordings with more speakers. Training with a set that follows the same proportion of speakers seen in CH1 (+SC 2-7\,spk) does not bring considerable gains. However, it should be noted that the training set is rather small and it is possible that learning to deal with more speakers requires a larger training set. 
Overall, the best results are obtained when using the 1-4 set for adaptation which is considerably larger than the other ones. This suggests that it would be beneficial to produce training data on-the-fly in order to encourage larger variability in the training set.

Table~\ref{tab:vad_loss} presents results when using the additional VAD loss described in Section~\ref{sc_multispk_vadloss} ($\alpha$$=$$0.2$) when adapting the SC 2\,spk model to 2-\textbf{4}\,spk (see Table~\ref{tab:multispeaker}) and doing fine-tuning. The use of the auxiliary loss allows for more even levels of missed and false alarm speech and this permits the fine-tuning to improve further, reaching the performance obtained when using the larger 1-4\, spk set.

\begin{table}[t]
    \caption{Error comparison on CH2 when using auxiliary VAD loss.}
    \label{tab:vad_loss}
    \setlength{\tabcolsep}{4pt} 
    \centering
    \begin{tabular}{@{}
                  l 
                  c |
                  S[table-format=2.2] |
                  S[table-format=2.2]
                  S[table-format=2.2]
                  S[table-format=2.2] 
                  @{}}
    \toprule
    System & $L_{VAD}$ & \multicolumn{1}{c}{DER} & \multicolumn{1}{c}{Miss} & \multicolumn{1}{c}{FA} & \multicolumn{1}{c}{Conf.} \\
    \midrule
    SC 2 spk + SC 2-\textbf{4} spk & \xmark & 17.09 & 7.32 & 4.12 & 5.66 \\
    \hspace{2.1cm}+ CH1 & \xmark & 16.27 & 9.55 & 2.55 & 4.16 \\
    \midrule
    SC 2 spk + SC 2-\textbf{4} spk & \cmark & 16.91 & 5.88 & 5.26 & 5.77 \\
    \hspace{2.1cm}+ CH1 & \cmark & 16.07 & 9.06 &  2.88  & 4.13 \\
    \bottomrule
  \end{tabular}
  \vspace*{-0.2cm}
\end{table}

\vspace{-0.2cm}

\section{Conclusions}
\vspace{-0.2cm}

Recently proposed simulated conversations as training data for EEND have shown remarkable performance with respect to the original simulated mixtures in 2-speaker telephone conversations. In this work, we extended the approach for conversations with more speakers and have shown that the same trend holds. We have also generated simulated conversations with different wide-band datasets in order to have models suited to non-telephone scenarios. However, the results were not conclusive showing that many challenges remain in order to generate adequate wide-band training data.

\clearpage

\bibliographystyle{IEEEbib}
\bibliography{refs}

\end{document}